\documentclass[10pt,showpacs,preprintnumbers,amsmath,amssymb]{revtex4}
\usepackage{units}
\usepackage{bbold,fourier}
\usepackage{graphicx}
\usepackage{dcolumn}
\usepackage{bm,bbm,mathtools,esvect}
\usepackage{slashed}

\begin{document}

\title{QUATERNIONIC $\;$ AHARONOV-BOHM $\;$ EFFECT}
\vspace{2cm}
\author{ SERGIO GIARDINO}
\email{p12@ubi.pt}
\affiliation{\qquad
\\ Departamento de F\'isica  \& Centro de Matem\'atica e Aplica\c c\~oes, Universidade da Beira Interior\\
Rua Marqu\^es D'\'Avila e Bolama 6200-001 Covilh\~a, Portugal}

\begin{abstract}
\noindent A quaternionic analog of the Aharonov-Bohm effect is developed without 
the usual anti-hermitian operators in quaternionic quantum mechanics (QQM).
A quaternionic phase links the solutions obtained to ordinary complex wave functions, 
and new theoretical studies and  experimental tests are possible for them.
\end{abstract}

\maketitle
\section{Introduction}%
Quaternions are non-commutative, generalized complex numbers. An arbitrary quaternion may be written in terms
of two complex variables, $z$ and $\zeta$, so that
\begin{equation}\label{I0}
q=z+\zeta j.
\end{equation}
(\ref{I0}) is the symplectic notation for quaternions and $j$ is a complex unit that does not commute with the complex unit $i$; 
in fact, $ij=-ji$.  There are many sources for the mathematics of quaternions, and we recommend \cite{Rocha:2013qtt}. A generalization of
quantum mechanics by introducing quaternions replacing complex numbers has been developed, and an introduction to the subject and
an encyclopedic list of references may be found at \cite{Adler:1995qqm}. Conversely, compared to complex quantum mechanics (CQM),
there are few explicit solutions for quaternionic quantum mechanics (QQM). For example, we have several results considering 
the quaternionic Schr\"odinger equation \cite{Davies:1989zza,Davies:1992oqq,Ducati:2001qo,Nishi:2002qd,DeLeo:2005bs,Ducati:2007wp}, 
and relativistic solutions obtained from the Dirac equation \cite{Davies:1990pm,DeLeo:2013xfa,DeLeo:2015hza,Giardino:2015iia} 
have also been obtained. 

Furthermore, extremely few analytic results of QQM have been used for experimental work, and a review of these  
can be found at \cite{Brumby:1996xf}. Quite recently, a new study has been carried out \cite{Walter:2016hct,Adler:2016eup},
but observable quaternionic effects have never been detected, despite these efforts. All these previous studies have been 
conducted considering that QQM should written using anti-hermitian operators, which obey $\mathcal{A}^\dagger=-\mathcal{A}$. In
this article, we propose a simpler QQM, where neither hermiticity nor anti-hermiticity are supposed. At this moment, we cannot
say whether this choice will lead to a consistent theory. However, we present some new results that may give us some 
hope that a consistent quaternionic theory may be built without imposing anti-hermiticity on its operators.
If we may develop several physically consistent results and examples, the mathematical formalism will
be developed from this set of results that may hopefully be considered in experimental tests. This is, in fact, the historical path
that have been followed by quantum mechanics. More recently, theoretical physics have become increasingly mathematical, including QQM.
This is not a problem, because the knowledge is never lost and it is a reference to new studies, including this one. However, we must 
always come back to the physical phenomena in order to ascertain what the mathematical formalism describes.

Thus, in this article, we write a quaternionic Schr\"odinger equation whose Hamiltonian operator has no defined hermiticity. 
This Hamiltonian has been obtained by defining a quaternionic momentum operator inspired by the electromagnetic field Hamiltonian.
Then we have simply made a transformation that can be schematically written as
\begin{equation}\label{I1}
 \bm\Pi=\bm p-\bm\alpha \qquad\to\qquad \bm\Pi=-\hbar\big(\bm\nabla-\bm\alpha i-\bm\beta j|i\big)
\end{equation}
where $\bm p$ is the momentum, 
$\bm\alpha$ is the real magnetic vector potential and $\bm\beta$ is an arbitrary complex vector potential without a
previous physical interpretation. We borrowed he notation of (\ref{I1}) from \cite{DeLeo:1991mi}, where the complex unity is 
multiplied on the right, namely
$(F|i)=F\,i\neq i\,F$. We use this momentum operator to obtain quaternionic wave functions $\Psi$
related to complex wave functions $\psi$ through a unit quaternion, so that
\begin{equation}\label{I2}
 \Psi=K\psi,
\end{equation}
where $\psi$ satisfies the time-dependent Schr\"odinger equation and $K$ is time-independent. We will see that there 
are non-trivial solutions that obey these
conditions, and if we interpret $\bm\beta$ as having a magnetic character, we will see that a particle in an Aharonov-Bohm experiment
will experience a force in a radial direction. This is a pure quaternionic result that may, in principle, be tested experimentally
and possibly explain the difference in the interference pattern observed in the Aharonov-Bohm (AB) effect. If these arguments are correct, we 
can see that the AB-effect is in fact quaternionic in nature, and then a new chapter in QQM may be written.

This article is organized as follows: in Section \ref{S1} we present a method for obtaining the novel quaternionic
wave functions, and in Section \ref{S2} we present general solutions. In section \ref{S4} we compare our 
results with the well-known AB-effect, and Section \ref{S5} rounds off the article with our conclusions.
\section{QUATERNIONIC QUANTUM  MECHANICS\label{S1}}%
Solving quaternionic differential equations is not a simple task, and many methods are available, several of them are reviewed in
\cite{Ducati:2001qo}. In QQM, the anti-hermitian condition is ubiquitous \cite{Adler:1995qqm}, and it is an important constraint to 
the solutions. As stated in the introduction, we will study solutions where the Hamiltonian operator does not satisfy the anti-hermiticity
constraint. However, this does not preclude quaternionic wave functions from being restrictive.  In the following sub-sections,
we use the non-anti-Hermitian operator to determine a class of quaternionic wave functions that will be suitable to describe the 
quaternionic AB-effect.

\subsection{COMPLEX AND QUATERNIONIC WAVE-FUNCTIONS \label{S11}}
Let us introduce a quaternionic wave function $\Psi$ in symplectic notation as
\begin{equation}\label{Psi}
 \Psi=\Big(\phi+\chi\, j\Big)\exp\left(-\frac{i\varepsilon t}{\hbar}\right),
\end{equation}
so that $\phi$ and $\chi$ are time-independent complex functions, the real quantity $\varepsilon$ is the energy, and the exponential 
is defined to be on the right hand side  of the time-independent quaternionic part of the wave function. 
The time-dependent Schr\"odinger equation is defined as
\begin{equation}\label{scheq}
\mathcal{H}\,\Psi=\hbar\,\partial_t\Psi i,
\end{equation}
where the complex unit $i$ is defined to be on the right hand side of $\partial_t\Psi$, as in \cite{DeLeo:1991mi}. Thus,
(\ref{Psi}) and (\ref{scheq}) permit us to obtain the usual time-independent Schr\"odinger equation
\begin{equation}\label{eigenval}
\mathcal{H}\,\Phi=\varepsilon \Phi,\qquad\mbox{where}\qquad\Phi=\phi+\chi\, j
\end{equation}
and the Hamiltonian operator is also quaternionic, so that
\begin{equation}\label{hamilt}
\mathcal{H}=H+L\,j.
\end{equation}
The ordinary Sch\"odinger equation, $H\psi=i\hbar\partial_t\psi$, is obtained when $L=\chi=0$.
If $\mathcal H$ were complex, then (\ref{eigenval}) would be valid for $\phi$ and $\chi$ independently, so that $\phi=\chi$
and the quaternionic solution would be physically equivalent to the complex one.  
$H$ is a hermitian operator, but $\mathcal{H}$ has no defined hermiticity because $L$ is arbitrary. The Hamiltonian (\ref{hamilt})
and the wave function (\ref{Psi}) are used to split (\ref{eigenval}) into two complex equations
\begin{equation}\label{eqs}
\big(H-\varepsilon\big)\phi-L\bar\chi=0\qquad\mbox{and}\qquad \big(H-\varepsilon\big)\chi+L\bar\phi=0,
\end{equation}
where $\bar\phi$ and $\bar\chi$ are complex conjugates and the a complex number $z$ obeys the rule $zj=j\bar z$. In order to decouple (\ref{eqs}),
we have two cases, classified according to the commutativity of $H$ and $L$. If $\,[H,\,L\big]=0$, then (\ref{eqs}) implies
\begin{equation}\label{A0}
 \Big[\big(H-\varepsilon\big)+L\bar L\Big]\Xi=0,\qquad\mbox{for}\qquad\Xi=\phi,\chi.
\end{equation}
(\ref{A0}) means that $\phi$ and $\chi$ are either equal or related by a complex constant, so that (\ref{Psi}) is related to the complex case
by a quaternionic constant. Another possibility is that the Hamiltonian does not commute with
$L$, where $\,[H,\,L\big]=\theta$ and $\theta$ is a complex parameter. Using this commutation relation, we get
\begin{equation}\label{coup2}
\Big[\big(H-\varepsilon\big)^2+L\bar L\Big]\phi-\theta\bar\chi=0\qquad\mbox{and}\qquad 
\Big[\big(H-\varepsilon\big)^2+L\bar L\Big]\chi+\theta\bar\phi=0. 
\end{equation}
Equations (\ref{coup2}) are obtained from (\ref{eqs}) by the replacement 
\begin{equation}
\big(H-\varepsilon\big)\to\Big[\big(H-\varepsilon\big)^2+L\bar L\Big]\qquad\mbox{and}\qquad L\to\theta.
\end{equation}
Thus, if $\,\big[\big(H-\varepsilon\big)^2+L\bar L\big]$ commutes with $\,\theta$, then $\,\phi$ and $\,\chi$ are either equal or
related. Conversely, if they do not commute,
then we repeat the procedure employed to obtain (\ref{coup2}) from (\ref{eqs}) and evaluate the commutativity of the new operators.
This analysis enables us to suppose that $\phi$ and $\chi$ must always be related in order to have a quaternionic solution for the
 Schr\"odinger equation.
In the next sub-section, we use this conclusion to  obtain physically non-trivial quaternionic solutions.
\subsection{CONSTRAINTS TO THE QUATERNIONIC WAVE-FUNCTION}
In the previous section, the dependence between the time-independent symplectic components of the wave function (\ref{Psi}) has been pointed out. 
We wonder whether we can use the complex
solution of (\ref{scheq}), namely the solution that is obtained for the limit within $L\to 0$ in (\ref{hamilt}). Let us name $\Psi$ the 
quaternionic solution and $\psi$ the complex solution, such that 
\begin{equation}\label{wfc}
\Psi=\Phi\,\exp\left(-\frac{i\varepsilon t}{\hbar}\right),
\qquad\mbox{and}\qquad
\psi=\phi\,\exp\left(-\frac{i E t}{\hbar}\right).
\end{equation}
Let us suppose that the time-independent wave functions are related according to
\begin{equation}\label{201}
\Phi=K\,\phi,\qquad\mbox{where}\qquad K=\cos\Theta\, e^{i\,\Gamma}+\sin\Theta\, e^{i\,\Omega}\,j,
\end{equation}
$\Gamma,\,\Theta$ and $\Omega$ are space-dependent real functions and of course $|K|=1$. We call (\ref{201}) the left quaternionic
solution; in the appendix we discuss the right quaternionic solution $\Phi=\phi K$. We notice that, within the complex 
limit of $K$, 
\[
\Theta\to 0,\qquad\mbox{we have}\qquad\Phi\to e^{i\,\Gamma}\phi,
\]
and consequently there is a geometric phase relating the solutions within the complex limit of $\Psi$. If the proposal (\ref{201})
is correct, then it generalizes the geometric phase of quantum mechanics. We know that the AB-effect is one of the simplest cases of 
geometric phases, and
then we will try to obtain a quaternionic analogue of the AB-effect. In order to achieve this objective, we need
a quaternionic momentum to build a quaternionic Hamiltonian. To achieve this aim, we calculate  the
gradient of (\ref{201}) and we get
\begin{equation}\label{grad0}
\Big[\bm\nabla-\big(\bm\nabla K\big)K^{-1}\Big]\Phi=K\,\bm\nabla\phi.
\end{equation}
Then we define the momentum operator $\bm\Pi$, namely 
\begin{equation}\label{pidef}
\bm\Pi\,\Phi=-\hbar\big[\bm\nabla-\big(\bm\nabla K\big)K^{-1}\big]\Phi\,i,
\end{equation}
and we observe that the complex unit $i$ must be multiplied on the right of $\Phi$. From (\ref{grad0}) and (\ref{pidef}) we get
the squared momentum,
\begin{equation}\label{a1}
\bm\Pi(K\,\phi)=-\hbar K\,\bm\nabla\phi\,i\qquad\Rightarrow\qquad \bm\Pi^2(K\,\phi)=-\hbar^2 K\bm\nabla^2\phi.
\end{equation}
Now, we want to solve the quaternionic Schr\"odinger equation
\begin{equation}\label{qse}
 \left(\frac{\bm\Pi^2}{2m}+V\right)\Psi=\hbar\,\partial_t\Psi\,i.
\end{equation}
If $V$ is a real potential, the complex Schr\"odinger equation is, of course,
\begin{equation}\label{202}
 \left(-\frac{\hbar^2}{2m}\bm\nabla^2+V\right)\psi=\hbar\,\partial_t\psi\,i.
\end{equation}
Using the time-independent equation $H\phi=E\phi$ and (\ref{a1}) in (\ref{qse}), we get
\begin{equation}
 \varepsilon=E,
\end{equation}
and then the quaternionic and complex wave functions related by (\ref{201})  
have the same energy for every real potential $V$. Now, we define an arbitrary quaternionic momentum given by
\begin{equation}\label{momento}
\bm\Pi\,\Phi=-\hbar\big(\bm\nabla-\bm Q\big)\Phi\,i,\qquad\mbox{where}\qquad \bm Q=\bm\alpha i+\bm\beta j
\end{equation}
with $\bm\alpha$ an arbitrary real vector and $\bm\beta$ an arbitrary complex vector. We notice that the probability density
is preserved in this model. Using (\ref{qse}) and (\ref{momento}), we get the continuity equation
\begin{equation}\label{A5}
\frac{\partial \rho}{\partial t}+ \bm\nabla\cdot \bm j=0,\qquad\mbox{where}\qquad
\rho=\Psi\Psi^*\qquad\mbox{and}\qquad
\bm j=\frac{1}{2m}\Big[\Psi^*\Pi\Psi+\big(\Psi^*\Pi\Psi\big)^*\Big].
\end{equation}
The conservation of the probability density ascertained by (\ref{A5}) is an important consistency test. Furthermore,
it is an example of non-anti-hermitian quaternionic quantum model. Quaternionic quantum mechanics has been developed using
anti-hermitian operators \cite{Adler:1995qqm}, and thus we have a counterexample showing that a consistent non-anti-hermitian 
model is possible. Substituting $\Phi$
in (\ref{qse}), and using (\ref{wfc}), (\ref{momento}) and (\ref{202}), we get
\begin{equation}\label{q_ctr}
\Big[\bm\nabla^2 K-\big(\bm{\nabla\cdot Q}\big)K-2\bm{Q\cdot\nabla}K+\bm{Q\cdot Q}\, K\Big]\phi+2\Big(\bm\nabla K-\bm Q K\Big)\bm{\cdot\nabla}\phi=0.
\end{equation}
(\ref{q_ctr}) generates equations for $\Theta,\,\Gamma,$ and $\,\Omega$ that permit us to determine the quaternionic wave function 
and its spectrum. Solving these constraints may be a difficult task, as we see from the derivatives of $K$, which
we write schematically as
\begin{equation}\label{KK}
 \bm\nabla K=\bm p e^{i\Gamma}+\bm q e^{i\Omega}j\qquad\mbox{and}\qquad \nabla^2 K=u e^{i\Gamma}+ v e^{i\Omega}j,
\end{equation}
where
\begin{align}
& \bm p=-\sin\Theta\,\bm\nabla\Theta+i\cos\Theta\,\bm\nabla\Gamma,\qquad
\bm q= \cos\Theta\,\bm\nabla\Theta+i\,\sin\Theta\,\bm\nabla\Omega,\nonumber\\
& u= -\cos\Theta\,\Big(\,\big|\bm\nabla\Gamma\big|^2+\big|\bm\nabla\Theta\big|^2\,\Big)-\sin\Theta\,\nabla^2\Theta
+i\Big(\cos\Theta\,\nabla^2\Gamma-2\sin\Theta\,\bm\nabla\Gamma\bm{\cdot\nabla}\Theta\Big)\label{pquv}\\
 & v=-\sin\Theta\,\Big(\,\big|\bm\nabla\Omega\big|^2+\big|\bm\nabla\Theta\big|^2\,\Big)+\cos\Theta\,\nabla^2\Theta
+i\Big(\sin\Theta\,\nabla^2\Omega+2\cos\Theta\,\bm\nabla\Omega\bm{\cdot\nabla}\Theta\Big).\nonumber
\end{align}
In order to obtain solutions, we use (\ref{KK}) to split (\ref{q_ctr}) into pure complex and pure quaternionic equations. 
The complex equation reads
\begin{equation}\label{cqe}
 \Big[\Big(u-i\cos\Theta\,\bm{\nabla\cdot\alpha}-2i\bm{\alpha\cdot p}+\cos\Theta\,\bm{Q\cdot Q}\Big) e^{i\Gamma} 
+\Big(\sin\Theta\,\bm{\nabla\cdot\beta}+2\bm{\beta\cdot}\bar{\bm q}\Big)e^{-i\Omega}\Big]\phi+
2\Big[\Big(\bm p-i\cos\Theta\,\bm\alpha\Big)e^{i\Gamma}+\sin\Theta\,\bm\beta e^{-i\Omega}\Big]\bm{\cdot\nabla}\phi=0,
\end{equation}
and the quaternionic term is
\begin{equation}\label{qqe}
 \Big[\Big(\bar v+i\sin\Theta\,\bm{\nabla\cdot\alpha}+2i\bm{\alpha\cdot}\bm{\bar q}+\sin\Theta\,\bm{Q\cdot Q}\Big) e^{-i\Omega} 
-\Big(\cos\Theta\,\bm{\nabla\cdot}\bm{\bar\beta}+2\bm{\bar\beta\cdot p}\Big)e^{i\Gamma}\Big]\phi+
2\Big[\Big(\bm{\bar q}+i\sin\Theta\,\bm\alpha\Big)e^{-i\Omega}-\cos\Theta\,\bm{\bar\beta} e^{i\Gamma}\Big]\bm{\cdot\nabla}\phi=0.
\end{equation}
Equations (\ref{cqe}) and (\ref{qqe}) are constraints that $\bm\alpha,\,\bm\beta,\,\Gamma,\,\Theta$ and $\Omega$ satisfy for each complex
solution $\phi$. The equations are not necessarily identical, and may differ from a complex factor, that can be either a constant or 
a function. Let us call $\lambda$ the factor that relates (\ref{cqe}) and (\ref{qqe}), then the coefficients for $\phi$ 
must satisfy
\begin{align}
&u-i\cos\Theta\,\bm{\nabla\cdot\alpha}-2i\bm{\alpha\cdot p}+\cos\Theta\,\bm{Q\cdot Q}=
-\lambda\Big(\cos\Theta\,\bm{\nabla\cdot}\bm{\bar\beta}+2\bm{\bar\beta\cdot p}\Big)\label{C1}\\
&\bar v+i\sin\Theta\,\bm{\nabla\cdot\alpha}+2i\bm{\alpha\cdot}\bm{\bar q}+\sin\Theta\,\bm{Q\cdot Q}=
\frac{1}{\lambda}\Big(\sin\Theta\,\bm{\nabla\cdot\beta}+2\bm{\beta\cdot}\bar{\bm q}\Big).\label{C2}
\end{align}
Accordingly, the coefficients for $\bm\nabla\phi$ are related by
\begin{equation}
\bm p-i\cos\Theta\,\bm\alpha=-\lambda\cos\Theta\,\bm{\bar\beta}\qquad\mbox{and}\qquad
\bm{\bar q}+i\sin\Theta\,\bm\alpha=\frac{1}{\lambda}\sin\Theta\,\bm\beta.\label{Cnabla}
\end{equation}
On the other hand, from (\ref{pquv}) and (\ref{Cnabla}) we get the vectors
\begin{equation}\label{C4}
\bm\alpha=\frac{1}{1+|\lambda|^2}\Big[\bm\nabla\Gamma+|\lambda|^2\bm\nabla\Omega+i\Big(\tan\Theta-|\lambda|^2\cot\Theta\Big)\bm\nabla\Theta\Big],\qquad
\bm\beta=\frac{\lambda}{1+|\lambda|^2}\left[\frac{2}{\sin 2\Theta}\bm\nabla\Theta+i\bm\nabla\big(\Gamma-\Omega\big)\right],
\end{equation}
so that the reality of $\bm\alpha$ is guaranteed by the constraint
\begin{equation}\label{C5}
\left(\tan^2\Theta-|\lambda|^2\right)\bm\nabla\Theta=0.
\end{equation}
The above vectors and constraints imply that it is possible to build up a generalized momentum $\bm\Pi$ in order to solve 
(\ref{qse}). Let us then consider whether we can find wave functions that satisfy (\ref{qse}). Using (\ref{cqe}) and (\ref{C1}-\ref{Cnabla}),
we get
\begin{equation}\label{C6}
 \Big[\Big(u-i\cos\Theta\,\bm{\nabla\cdot\alpha}-2i\bm{\alpha\cdot p}-\cos\Theta\,\eta\Big) e^{i\Gamma} 
+\Big(\bar v+i\sin\Theta\,\bm{\nabla\cdot\alpha}+2i\bm{\alpha\cdot}\bm{\bar q}-\sin\Theta\,\eta\Big)\lambda e^{-i\Omega}\Big]\phi+
2\Big[-\cos\Theta\,\bm{\bar\beta}\lambda e^{i\Gamma}+\sin\Theta\,\bm\beta e^{-i\Omega}\Big]\bm{\cdot\nabla}\phi=0.
\end{equation}
Thus, from (\ref{KK}-\ref{pquv}) and (\ref{C4}) we rewrite (\ref{C6}) as
\begin{align}\nonumber
&\left[\frac{2i|\lambda|}{\big(1+|\lambda|^2\,\big)^2}\Big( \cos\Theta\,e^{i\Gamma}-\lambda\sin\Theta\,e^{-i\Omega}\Big)\bm\nabla|\lambda|\bm{\cdot\nabla}(\Gamma-\Omega)\,
-\frac{2i}{1+|\lambda|^2}\Big(|\lambda|^2 \sin\Theta\,e^{i\Gamma}-\lambda\cos\Theta\,e^{-i\Omega}\Big)\bm\nabla(\Gamma-\Omega)\bm{\cdot\nabla}\Theta\right.-
\\ \nonumber
&\left. -\,\Big( \cos\Theta\,e^{i\Gamma}+\lambda\sin\Theta\,e^{-i\Omega}\Big)
\left(1+\frac{4}{\sin^2 2\Theta}\frac{|\lambda|^2}{\big(1+|\lambda|^2\,\big)^2}\right)|\bm\nabla\Theta|^2\,
-\,\Big( \sin\Theta\,e^{i\Gamma}-\lambda\cos\Theta\,e^{-i\Omega}\Big)\nabla^2\Theta\right]\phi+\,
\\ \nonumber
&+\frac{1}{1+|\lambda|^2}\Big(|\lambda|^2 \cos\Theta\,e^{i\Gamma}+\lambda\sin\Theta\,e^{-i\Omega}\Big)
\Big[\Big(i\nabla^2(\Gamma-\Omega)\,-|\bm\nabla(\Gamma-\Omega)|^2\Big)\phi\,+2i\bm\nabla(\Gamma-\Omega)\bm{\cdot\nabla}\phi\Big]-
\\ \label{C7}
&
-\frac{2}{1+|\lambda|^2}\left(|\lambda|^2\cos\Theta e^{i\Gamma}-\lambda\sin\Theta e^{-i\Omega}\right)
\frac{\bm\nabla\Theta\bm{\cdot\nabla}\phi}{\sin\Theta\cos\Theta}=0.
\end{align}
The vector potentials $\bm\alpha$ and $\bm\beta$, the constraints (\ref{C1}-\ref{Cnabla}) and the equation (\ref{C7}) is
the set of conditions that should be satisfied so that (\ref{wfc}-\ref{201}) is the solution of (\ref{qse}). 
In the following sections we study the solutions that can be obtained from these conditions.

\section{\label{S2}THE WAVE-FUNCTIONS}
Let us first consider the $\bm\nabla\Theta=0$ case, where (\ref{C7}) simplifies to
\begin{equation}\label{CC1}
\frac{|\lambda|^2 \cos\Theta\,e^{i\Gamma}+\lambda\sin\Theta\,e^{-i\Omega}}{1+|\lambda|^2}
\Big[\Big(i\nabla^2(\Gamma-\Omega)\,-|\bm\nabla(\Gamma-\Omega)|^2\Big)\phi\,+2i\bm\nabla(\Gamma-\Omega)\bm{\cdot\nabla}\phi\Big]=0.
\end{equation}
In order not to constrain the complex wave function, we impose conditions to get two cases. Now we will examine the first.
\subsection{SOLUTIONS FOR $\bm\nabla\Theta=0$ AND $\bm\nabla(\Gamma-\Omega)=0$}
In this case, the quaternionic solution comprises
\begin{equation}\label{teta0}
\Psi=e^{i\Omega}\,L\,\phi\,e^{-iEt/\hbar},\qquad\bm\alpha=\bm\nabla\Omega,\qquad\bm\beta=0,
\end{equation}
where $L$ is an arbitrary constant unitary quaternion. This is the simplest quaternionic solution
that is related to a complex solution through a geometric phase. It is also the simplest counterpart of the Aharonov-Bohm
effect that may be obtained in quaternionic quantum mechanics. Its interpretation is similar to the complex case, but in 
this case the phase is quaternionic and consequently non-commutative, and this is its physical novelty that can be researched 
experimentally. It reduces immediately 
to the complex case for $L=1$, and it is thus the simplest quaternionic generalization of the AB-effect.
\subsection{\label{ABQ}SOLUTIONS FOR $\bm\nabla\Theta=0$ AND $|\lambda|^2=\tan^2\Theta $}
In this case, (\ref{CC1}) is satisfied for
\begin{equation}\label{DD2}
\lambda=-\tan\Theta e^{i(\Gamma+\Omega)}.
\end{equation}
The condition implies that the quaternionic solution comprises
\begin{equation}\label{DD3}
\Phi=\left(\cos\Theta e^{i\Gamma}+\sin\Theta e^{i\Omega}j\right)\,\phi\,e^{-iEt/\hbar},
\qquad\bm\alpha=\cos^2\Theta\,\bm\nabla\Gamma+\sin^2\Theta\,\bm\nabla\Omega,\qquad
\bm\beta=-i\sin\Theta\cos\Theta\, e^{i(\Gamma+\Omega)}\bm\nabla\big(\Gamma-\Omega\big). 
\end{equation}
This solution reduces to the previous one if $\bm\nabla(\Gamma-\Omega)=0$, and then we see that it is another 
example of non-commutative quaternionic AB-effect. Furthermore, the fields generated by the vector potentials are interesting 
because $\bm{\nabla\times\beta}\neq 0$. Accordingly,
\begin{equation}
\bm{\nabla\times\alpha}=\bm 0\qquad\mbox{and}
\qquad\bm{\nabla\times\beta}=-\sin2\Theta\, e^{i(\Gamma+\Omega)}\,\bm\nabla\Gamma\bm{\times\nabla}\Omega. 
\end{equation}
Then, as in the previous case, the geometric phase is non-commutative. Furthermore, we have two non-trivial fields 
encoded in $\bm{\nabla\times\beta}$. Hence there is a big difference in relation to the usual complex AB-effect. In the 
quaternionic case, there are real fields in the region where the wave function is defined, so that the particle may
interact with the field in some manner. This interaction may explain change in the interference pattern observed in the AB-effect.
Additionally, some kind of non-commutativity must be observed, and this is the crucial point that must be stressed when researching 
quaternionic effects.

\subsection{THE $\bm\nabla\Theta\neq 0$ CASE}
Constraint (\ref{C5}) enables us to impose 
\begin{equation} \label{D0}
\lambda=\tan\Theta e^{i\vartheta}.
\end{equation}
Now, we consider the effect of this choice on the constraints (\ref{C1}-\ref{Cnabla}) and (\ref{C7}). Constraints
(\ref{Cnabla}) are trivially satisfied. Conversely, the complex constraints (\ref{C1}-\ref{C2}) generate four 
real conditions. One of these constraints implies that $|\bm\nabla\Theta|=0$, which is a contradiction, and then there
are no solutions in this case.
\section{\label{S4} One example }
We compare our results with the usual Aharonov-Bohm effect, where a a long solenoid of radius $R$ generates a magnetic field $\bm B$. 
Defining $r$ as a cylindrical radial distance, we get a zero magnetic field for $r>R$ and a constant magnetic field with intensity $B$ 
for $r<R$. Outside the solenoid, although $B=0$, the magnetic vector potential is finite and given by
\begin{equation}\label{E0}
 \bm A=\frac{\varPhi}{2\pi r}\bm{\hat\varphi},
\end{equation}
where  $\varPhi=\pi R^2B$ is the magnetic flux and $\bm{\hat\varphi}$ is the azimuthal angle direction. Let us consider the case described in
section \ref{ABQ}, where $\bm\nabla\Theta=0$ and $|\lambda|^2=\tan^2\Theta$, and the decomposition of $\bm{\hat\varphi}$ onto Cartesian 
coordinates
\begin{equation}\label{E1}
\bm{\hat\varphi}=-\sin\varphi\,\hat{\bm x}+\cos\varphi\,\hat{\bm y}.
\end{equation}
Remembering that $\bm\alpha=(q/\hbar)\bm A$, and using (\ref{DD3}), we may choose  
\begin{equation}\label{E2}
\bm\nabla\Gamma=-|\bm\alpha|\frac{\sin\varphi}{\cos^2\Theta}\hat{\bm x}\qquad\mbox{and}\qquad
\bm\nabla\Omega=|\bm\alpha|\frac{\cos\varphi}{\sin^2\Theta}\hat{\bm y}
\end{equation}
where $\bm\alpha=q/\hbar\bm A$. (\ref{DD3}) enables us to get the complex field
\begin{equation}\label{E3}
 \bm{\nabla\times\beta}=2|\bm\alpha|^2\frac{\sin2\varphi}{\sin 2\Theta}e^{i(\Gamma+\Omega)}\hat{\bm z}.
\end{equation}
The fact that (\ref{E3}) is complex is not necessarily a problem. The magnetic field also comes from an imaginary component of
(\ref{momento}). Conversely, the physical nature of the field generated by the potential $\bm\beta$ is not clear. If we suppose that
(\ref{E3}) has a magnetic character, then an electrically charged particle that describes a path on the azimuthal direction 
$\bm{\hat\varphi}$ will suffer a force over the radial direction given by the Lorenz expression. This may explain the interference
pattern of the AB-effect. Conversely, the non-commutativity of the wave function must also be ascertained in order to
verify the quaternionic character of the phenomenon. Then, non-commutativity and a quantitative correspondence between the interference
pattern and the radial quaternionic force are the phenomenona that must be experimentally researched in order to verify
the quaternionic AB-effect.

\section{conclusion\label{S5}}
In this article we obtained a quaternionic solution for the AB-effect. This solution was not obtained according to
the canonical anti-hermitian quaternionic quantum mechanics, although, a physically
meaningful solution was obtained. This fact opens a discussion about the mathematical consistence of non-anti-hermitian
QQM, and there is a necessity for results. Expected values, commutators, the Heisenberg formalism and the Ehrenfest
theorem are several examples of  the consistency tests that must be executer in order to understand whether a non-hermitian QQM
makes sense. 

If we suppose that a non-hermitian QQM is consistent, we have further interesting problems related to the physical solution described
in this paper. The AB-effect is an example of geometric phase, and a quaternionic case opens the possibility that a quaternionic
geometric phase must exist. The geometric phase of anti-hermitian QQM is already described \cite{Adler:1995qqm,Maia:2001dm}, but a
non-anti-Hermitian counterpart has not been developed.

Finally, we have the possible experiments that may be conducted in order to verify the predictions of the article. The possibility of
describing the AB-effect through a non-commutative quaternionic wave function seems to be an interesting theme for experimentalists,
and it is the only way to determine whether QQM is physically meaningful. Thus, we expect this article to attract interest in QQM
 because it raises interesting questions regarding conceptual issues, mathematical consistency and experimental tests.

\paragraph*{ACKNOWLEDGEMENTS}
This research received financial a grant from CNPq under number 206383/2014-2. The author is also grateful for the hospitality of 
Professor Paulo Vargas Moniz and the Center for Mathematics and Applications of the Beira Interior University during the academic
year 2015/2016.

\appendix
\section{Right quaternionic wave equation\label{A}}
In this article, we have considered the left quaternionic wave function defined in (\ref{201}). In this appendix, we present
the results for the right quaternionic wave function
\begin{equation}\label{A1}
\Phi=\phi\,K,\qquad\mbox{where}\qquad K=\cos\Theta\, e^{i\,\Gamma}+\sin\Theta\, e^{i\,\Omega}\,j,
\end{equation}
 $\Gamma,\,\Theta$ and $\Omega$ are space-dependent real functions and $|K|=1$. We limit ourselves to present the most important
results concerning the left quaternionic wave function discussed in section \ref{S1}. The more general case, where a quaternionic
wave function is $\Phi=K\,\phi\,L$, with $K$ and $L$ unit quaternions, is left for future work. Using (\ref{A1}) in (\ref{qse}),
we get
\begin{equation}\label{A2}
 \phi\,\nabla^2K-2\bm{Q\cdot}\big(\phi\bm\nabla K\big)-\big(\bm{\nabla\cdot Q}-\eta\big)\phi\,K+2\Big[\bm\nabla\phi\bm{\cdot\nabla}K
-\big(\bm{Q\cdot\nabla}\phi\big) K\Big]=0,
\end{equation}
where $\eta=\bm{Q\cdot Q}$. Using the definition of $K$ (\ref{A1}) and its derivatives (\ref{KK}), the pure complex
part of (\ref{A2}
\begin{equation}\label{A3}
\Big(u-i\cos\Theta\,\bm{\nabla\cdot\alpha}-2i\bm{\alpha\cdot p}+\cos\Theta\,\eta\Big)\phi\,e^{i\Gamma} 
+\Big(\sin\Theta\,\bm{\nabla\cdot\beta}+2\bm{\beta\cdot}\bar{\bm q}\Big)\bar\phi e^{-i\Omega}+
2\Big[\Big(\bm p-i\cos\Theta\,\bm\alpha\Big)\bm{\cdot\nabla}\phi\,e^{i\Gamma}+\sin\Theta\,\bm\beta\bm{\cdot\nabla}\bar\phi\,e^{-i\Omega}\Big]=0,
\end{equation}
while the pure quaternionic term of (\ref{A2}) is
\begin{equation}\label{A4}
\Big(\bar v+i\sin\Theta\,\bm{\nabla\cdot\alpha}+2i\bm{\alpha\cdot}\bm{\bar q}+\sin\Theta\,\eta\Big)\,\bar\phi e^{-i\Omega} 
-\Big(\cos\Theta\,\bm{\nabla\cdot}\bm{\bar\beta}+2\bm{\bar\beta\cdot p}\Big)\,\phi\,e^{i\Gamma}+
2\Big[\Big(\bm{\bar q}+i\sin\Theta\,\bm\alpha\Big){\cdot\nabla}\bar\phi\, e^{-i\Omega}
-\cos\Theta\,\bm{\bar\beta}\bm{\cdot\nabla}\phi\, e^{i\Gamma}\Big]=0.
\end{equation}
We see that (\ref{A3}) and (\ref{A4}) generate conditions identical to (\ref{C1}) and (\ref{C2}), and then we conclude that
the left quaternionic wave function and the right and left quaternionic solutions have common general potential vectors (\ref{C4}) 
and common constraints (\ref{C5}), and then the physical content of right quaternionic wave functions is equivalent to that of
left quaternionic wave functions.
%
%
%
%

\bibliographystyle{unsrt} 
\bibliography{bib_weq}

\begin{thebibliography}{10}

\bibitem{Rocha:2013qtt}
\texttt{J. Vaz; R. da Rocha }.
\newblock {``An Introduction to Clifford Algebras and Spinors''}.
\newblock Oxford University Press (2016).

\bibitem{Adler:1995qqm}
\texttt{S. L. Adler}.
\newblock {``Quaternionic Quantum Mechanics and Quantum Fields''}.
\newblock Oxford University Press (1995).

\bibitem{Davies:1989zza}
{\tt A. J. Davies; B. H. J. McKellar}.
\newblock {``Nonrelativistic quaternionic quantum mechanics in one
  dimension''}.
\newblock {\em Phys. Rev.}, {\bf A40}:4209--4214, (1989).

\bibitem{Davies:1992oqq}
{\tt A. J. Davies; B. H. J. McKellar}.
\newblock {``Observability of quaternionic quantum mechanics''}.
\newblock {\em Phys. Rev.}, {\bf A46}:3671--3675, (1989).

\bibitem{Ducati:2001qo}
\texttt{S. De Leo; G. Ducati}.
\newblock {`` Quaternionic differential operators''}.
\newblock {\em J. Math. Phys}, {\bf 42}:2236--2265, (2001).

\bibitem{Nishi:2002qd}
\texttt{S. De Leo; G. Ducati}.
\newblock {`` Quaternionic potentials in non-relativistic quantum mechanics''}.
\newblock {\em J. Phys}, {\bf A35}:5411--5426, (2002).

\bibitem{DeLeo:2005bs}
\texttt{S. De Leo; G. Ducati}.
\newblock {`` Quaternionic bound states''}.
\newblock {\em J. Phys}, {\bf A35}:3443--3454, (2005).

\bibitem{Ducati:2007wp}
\texttt{S. De Leo; G. Ducati}.
\newblock {`` Quaternionic wave packets''}.
\newblock {\em J. Math. Phys}, {\bf 48}:052111--10, (2007).

\bibitem{Davies:1990pm}
\texttt{A. J. Davies}.
\newblock {``Quaternionic Dirac equation''}.
\newblock {\em Phys.Rev.}, {\bf D41}:2628--2630, (1990).

\bibitem{DeLeo:2013xfa}
\texttt{S. De Leo; S. Giardino}.
\newblock {``Dirac solutions for quaternionic potentials''}.
\newblock {\em J. Math. Phys.}, {\bf 55}:022301--10, (2014) {\tt
  arXiv:1311.6673[math-ph]}.

\bibitem{DeLeo:2015hza}
\texttt{S. De Leo; G. Ducati; S. Giardino}.
\newblock {``Quaternioninc Dirac Scattering''}.
\newblock {\em J. Phys. Math.}, {\bf 6}:1000130, (2015) {\tt
  arXiv:1505.01807[math-ph]}.

\bibitem{Giardino:2015iia}
\texttt{S. Giardino}.
\newblock {``Quaternionic particle in a relativistic box''}.
\newblock {\em Found. Phys.}, {\bf 46}(4):473--483, (2016){\tt
  arXiv:1504.00643[quant-ph]}.

\bibitem{Brumby:1996xf}
\texttt{S. P. Brumby; G. C. Joshi}.
\newblock {``Experimental status of quaternionic quantum mechanics''}.
\newblock {\em Chaos Solitons Fractals}, {\bf 7}:747--752, (1996) {\tt
  quant-ph/9610009}.

\bibitem{Walter:2016hct}
{\tt L. M. Procopio; L. A. Rozema; Z. J. Wong; D. R. Hamel; K. O'Brien; X.
  Zhang; B. Dakic; P. Walther}.
\newblock {``Experimental Test of Hyper-Complex Quantum Theories''}.
\newblock (2016) {\tt arXiv:1602.01624 [quant-ph] }.

\bibitem{Adler:2016eup}
{\tt S. L. Adler}.
\newblock {``Does the Peres experiment using photons test for hyper-complex
  (quaternionic) quantum theories?''}.
\newblock (2016) {\tt arXiv:1604.04950 [quant-ph]}.

\bibitem{DeLeo:1991mi}
\texttt{S. De Leo; P. Rotelli}.
\newblock {``The Quaternion scalar field''}.
\newblock {\em Phys.Rev.}, {\bf D45}:575--579, (1992).

\bibitem{Maia:2001dm}
\texttt{M. D. Maia; V. B. Bezerra}.
\newblock {``Geometric phase in quaternionic quantum mechanics''}.
\newblock {\em Int. J. Theor. Phys.}, {\bf 40}:1283--1294, (2001) {\tt
  hep-th/0107107}.

\end{thebibliography}

\end{document}